\documentclass[final,3p,times,numbers,sort&compress]{elsarticle}

\usepackage{graphicx}
\usepackage{amsmath}
\usepackage{amssymb}
\usepackage{bm}
\usepackage{color}
\usepackage{xcolor}
\usepackage{soul}    
\usepackage{caption}
\usepackage{subcaption}
\usepackage{resizegather}
\usepackage[utf8]{inputenc}
\usepackage{url}

\journal{Physica A}

\begin{document}

\begin{frontmatter}

\title{Multiple metastable states in an off-lattice Potts model}

\author[unab]{Constanza Farías\corref{cor1}}
\ead{m.fariasparra@uandresbello.edu}
\address[unab]{Departamento de F\'isica, Facultad de Ciencias Exactas, Universidad Andres Bello. Sazi\'e 2212, piso 7, Santiago, 8370136, Chile.}

\author[cchen,unab]{Sergio Davis}
\address[cchen]{Comisión Chilena de Energía Nuclear, Casilla 188-D, Santiago, Chile}

\cortext[cor1]{Corresponding author}

\begin{abstract}
The interactions between a group of components are commonly studied in several areas of science (social science, biology, material science, complex dynamical systems,
among others) using the methods of thermodynamics and statistical mechanics. In this work we study the properties of the recently proposed off-lattice, two-dimensional Potts model
[Eur. Phys. J. B \textbf{87}, 78 (2014)], originally motivated by the dynamics of agent opinions, and which is described by a Hamiltonian obtained by a maximum entropy inference
procedure. We performed microcanonical and canonical Monte Carlo simulations of the first-order phase transition in the model, revealing a caloric curve with metastable regions.
Furthermore, we report a ``switching'' behavior between multiple metastable states. We also note that the thermodynamics of the model has striking similarities with systems having
long-range interactions, even though the interactions are short-ranged.
\end{abstract}


\end{frontmatter}



\section{Introduction}

Spin models, such as the Ising model and its generalization, the Potts model~\cite{Wu1982}, have been proposed to describe correlation between individual sites, in different fields of Science
including condensed matter physics\cite{Louis2018}, neuroscience\cite{Naim2018}, social science\cite{Bisconti2015}, percolation theory\cite{Debenedetti2001}, among many others.

As is the case with several types of interacting systems presenting phase transitions, it is well known that in spin systems the phenomenon of metastability can be
observed~\cite{Moreno2018, Nishino2018}. A metastable state is a quasi-stable state of a dynamical system, where the macroscopic observables reach values in a steady state
which, nevertheless, has a finite lifetime~\cite{Shell2015}. In the study of metastability is common the use of the microcanonical ensemble, in which we can observe precisely these
forbidden states that cannot be reached within the canonical ensemble. Typical examples of metastable states are superheated solids and supercooled liquids, and in both cases the
state appears at some temperatures where interesting kinetics appears. However, metastability is also observed outside the scope of thermodynamic systems, for instance in
complex dynamical systems~\cite{Zeng2020}. In these systems, multiple metastable states are extremely sensitive to noise or perturbations, and the initial conditions are responsible
for coexistence of different attractors~\cite{Feudel2008}. Moreover multistability plays an important role in some of the basic processes of life, and might even account for the
\emph{maintenance of phenotypic differences in the absence of genetic or environmental distinctions}~\cite{Laurent1999}.

In this work, we study a opinion model in social groups that takes into account how internal and external opinions may agree with each other~\cite{Davis2014}.
In this model, unlike most studies on spin Hamiltonians where the spins are fixed on a lattice, we consider free-moving spins, i.e. not attached to an underlying lattice. We
incorporate interactions between the agents, described by a Hamiltonian that is a variant of the two-dimensional Potts model, in which spins are free-moving particles delimited
by a square box of length $L$, interacting within a radius $R_c$.

This paper is organized as follows. In Section 2, we define the model and its parameters, and give details on the Monte Carlo methodology used to study it, while Section 3 presents
the results of a canonical and microcanonical study of first-order phase transitions in the model, together with some dynamical aspects such as multistability. Finally we close with
some concluding remarks in Section 4.

\section{Computational Methods}
\label{sec:headings}

Before describing the main topic of this section, it is important to contextualize it to the social world and the interactions between individuals who might share certain opinions
on a particular topic. Consider a group of $N$ individuals who interact with each other according to their $Q$ possible opinions on a subject; normally human
beings can behave socially in one of two ways, namely giving an opinion according to what they truly think, or
instead manifest an opinion which is not consistent with what they actually believe, in order to ``fit in'' better with a social group. In this work we consider individuals who are mostly
consequent with their thoughts.

We will consider for our model that the opinions that agents manifest are the ``spins'' $S_i$, while in turn the agents have internal beliefs $B_i$, which may or may not coincide with the
manifest opinion $S_i$ depending on the strength of the ``consequence'' parameter $C$. In addition, the parameter $J > 0$ will play the role of an
exchange coupling, which in this work can be understood as the degree of agreement among neighbors.
These parameters are such that $C>J$ for the case when these agents mostly coincide their internal beliefs with the manifest opinion regardless of the external influence. On the contrary, in the case where $C<J$, internal and external opinions do not necessarily coincide, indicating that the agents are influenced by the opinions
of others against their own beliefs. We will see how these two parameters arise, in the following.

The starting point of our model is the probability $p_C$ of having a consistent opinion, together with the probability $p_A$ of two agents to agree with each other given that
they are closer than the radius $R_c$. These can be written in the form of expectation constraints as
\begin{align}
P(S_i = B_i | I) & = p_C \quad \forall\;i, \\
P(S_i = S_j | r_{ij} < R_c, I) & = p_A \quad \forall\;i,j,
\end{align}
respectively. After a maximum entropy inference~\cite{Jaynes1957} procedure as in Ref.~\cite{Castellana2014} (see the Appendix for more details), we are led to the Hamiltonian of
Ref.\cite{Davis2014}, which is given in terms of the $S_i$ and $B_i$ as follows,
\begin{equation}
H = -\frac{1}{2}J\sum_{i=1}^{N}\sum_{<j\neq i>} \Big<s_{i},s_{j}\Big> + \frac{1}{2}\sum_{i=1}^{N}\sum_{<j\neq i>}R - C\sum_{i=1} \Big<s_{i},B_{i}\Big>,
\end{equation}
where $$\big<a,b\big> := 2\delta(a,b)-1$$ and $\delta(a,b)$ is the Kronecker delta. Note that the sum over $\big<j\neq i\big>$ term only considers the
pairs within the radius $R_c$, which represents the extent of the influence of agents upon others.

As in Ref.~\cite{Davis2014}, we are not interested in the effect of overcrowding so accordingly, the second term involving the $R$ parameter will be neglected. This is equivalent
to setting $p_A = \frac{1}{2}$, as shown in the Appendix. So in this way and after some rearrangements,
we obtain the final form of our Hamiltonian as
\begin{equation}
\label{eq_ham}
\frac{H}{C}= -\frac{1}{2}\frac{J}{C}\sum_{i=1}^{N}\sum_{<j\neq i>} \delta(s_i,s_j)\Theta(Rc-r_{ij}) - 2\sum_{i=1}\delta(s_{i},B_{i}).
\end{equation}

\noindent
For clarity, we will take $C$=1 and express the results in terms of the ratio $J/C$, while also using $C$ as the unit of energy. A given value of $J/C$ can be understood as a
degree of tendency to belong to a group by maintaining or changing their opinion, depending on its value. In this case we consider a positive value of $J/C < 1$, which means an
agent will tend to remain firm with his inner opinion and would tend to leave a group.

Please also note that the Hamiltonian as given by Eq. \ref{eq_ham} involves only the configurational degrees of freedom $S_i$ and $\bm{r}_i$, and therefore represents the
potential energy of the system. Accordingly, in the following we will denote this part as $\Phi$. Furthermore, potential energy in this model is higher if pairs of agents with the same
manifest opinion are farther than $R_c$ due to the interaction parameter $J$, and also higher if the agents manifest a different opinion than their internal belief, because of
the consequence parameter $C$.


\subsection{Monte Carlo Metropolis algorithm}

In order to determine the properties of this model, we have performed microcanonical and canonical ensemble
simulations using in the former case Ray’s version of the Monte Carlo Metropolis algorithm \cite{Ray1991}.

For the canonical ensemble, as is well known\cite{Tobochnik2008} the acceptance probability is given by
\begin{equation}
p_\text{acc} = \min\big ( 1, \exp{(-\beta \Delta \phi)} \Big )
\end{equation}
while in the microcanonical ensemble, it is given by
\begin{equation}
p_\text{acc} = \min\Big(1,\Big[\frac{E-\Phi'}{E-\Phi}\Big]^{\frac{dN}{2}-1}\Big)
\end{equation}

\noindent
where $d$ is the dimension of the system, and in our case $d=2$. Finally we obtain
\begin{equation}
p_\text{acc} = \min\Big(1,\Big[\frac{E-\Phi'}{E-\Phi}\Big]^{N-1}\Big).
\end{equation}

The Monte Carlo trial moves used for this work were considered under two conditions, regarding a probability $p_s$ for particles to change spin, or otherwise, move and interact with other agents. More precisely, if a uniform random
variable $x \in [0, 1)$ is such that, $x<p_s$ then a previously selected agent appears to ``change its mind'', that is, changes its spin. If, on the other hand,  $x \geq p_s$
the selected agent does not change opinion. Instead, it changes position and, in doing so may interact with other agents. In our work we used a value $p_s$=0.1.

\noindent
The temperature in the microcanonical ensemble $T(E)$ is obtained from the configurations by the relation~\cite{Carignano2002}
\begin{equation}
\frac{1}{T(E)} := \Big<\frac{dN}{2(E-\Phi)}\Big>_E = \Big<\frac{N}{(E-\Phi)}\Big>_E.
\end{equation}


\section{Results}

As was shown in Ref.\cite{Davis2014}, the model has both second-order and first-order phase transitions depending on the values of $\rho := R_c/L$ and $J/C$. In this study we have
chosen to focus our attention on the first-order phase transitions and the presence of metastable states. Therefore, for our simulations using the Monte Carlo Metropolis algorithm we
have used the two-dimensional $Q$=2 Potts model case, with a particle number $N$=100, where the values of parameters used are $\rho$ = 0.06, $J/C$ = 0.6 and $R_c$ = 6.708, in all cases
which are different as in \cite{Davis2014}, in the same way we made this simulation considering a square box of side $L$=111.8 units, which is proportional to $N$.

\subsection{Canonical and microcanonical thermodynamics of the model}

\begin{figure}[h]
  \centering
  \includegraphics[scale=0.55]{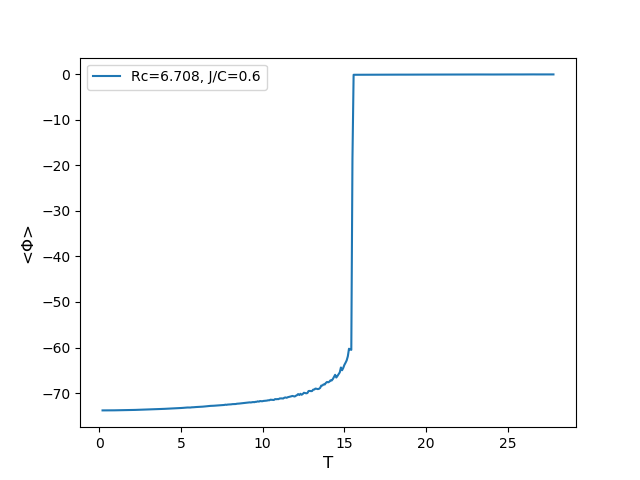}
  \caption{Caloric curve for the canonical ensemble, where a first-order phase transition can be seen. The y-axis show the expectation value of Potential Energy $\big<\Phi\big>$.}
  \label{fig_canonical}   
\end{figure}

\begin{figure}[h!]
  \centering
  \includegraphics[scale=0.55]{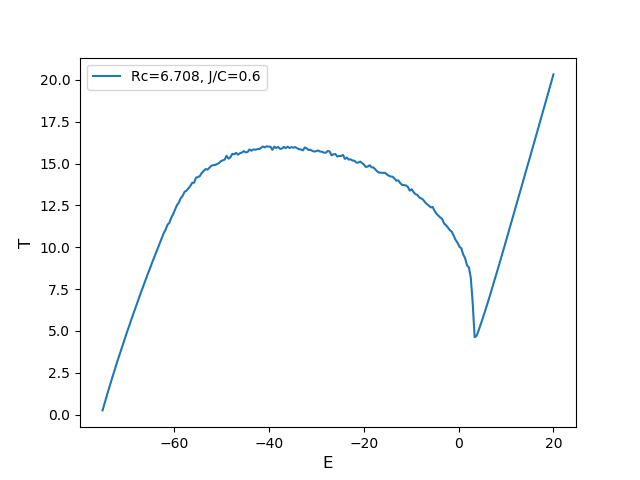}
  \caption{Caloric curve for the microcanonical ensemble, where a first-order phase transition, together with a region of metastability, can be seen.}
  \label{fig_microcanonical} 
\end{figure}

Fig.~\ref{fig_canonical} shows the canonical caloric curve $\big<\Phi\big>_\beta$ versus $T$, in which at low energies (approaching $\big<\Phi\big>$ = -60 from below), small perturbations can be seen due to being close to the critical point $T_c$ = 15.5 where the phase transition occurs. Similarly,
Fig.~\ref{fig_microcanonical} shows the microcanonical caloric curve $\big<T\big>_E$ versus $E$ where this kind of perturbation is seen in greater detail, as well as the phase
transition that in the caloric curve of the canonical ensemble is an abrupt vertical curve. The microcanonical caloric curve strikingly resembles those of systems with
long-range interactions~\cite{Casetti2010,Casetti2012}, despite the fact that our model only consider direct interactions within the cutoff $R_c$.

\begin{figure}[h!]
  \centering
  \includegraphics[scale=0.55]{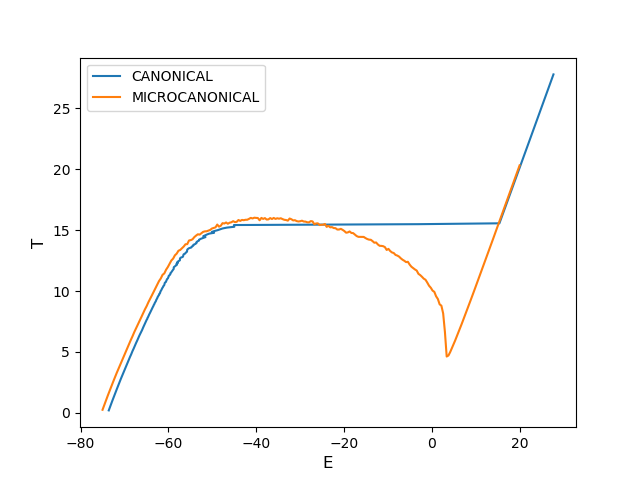}
  \caption{Cross curves between Microcanonical and Canonical Ensemble}
  \label{fig_mix}
\end{figure}

In Fig.~\ref{fig_mix} the crossing curves represent the microcanonical caloric curve of Fig.\ref{fig_microcanonical} superimposed with the canonical caloric curve
of Fig.\ref{fig_canonical} but where in the latter we have added the kinetic energy $\big<K\big>_\beta = N/\beta$.
We can observe an interesting situation, namely the presence of metastable states between the equilibrium, stable branches (straight lines) below $E$ = -60 and above $E$ = 20. A large metastable region appears  clearly between the intersections of the canonical and microcanonical curves at $E$=-25.6525 and $E$=15.2448.

\subsection{Metastable states}

%
%
%
\begin{figure}[h!]
         \includegraphics[width=0.35\textwidth]{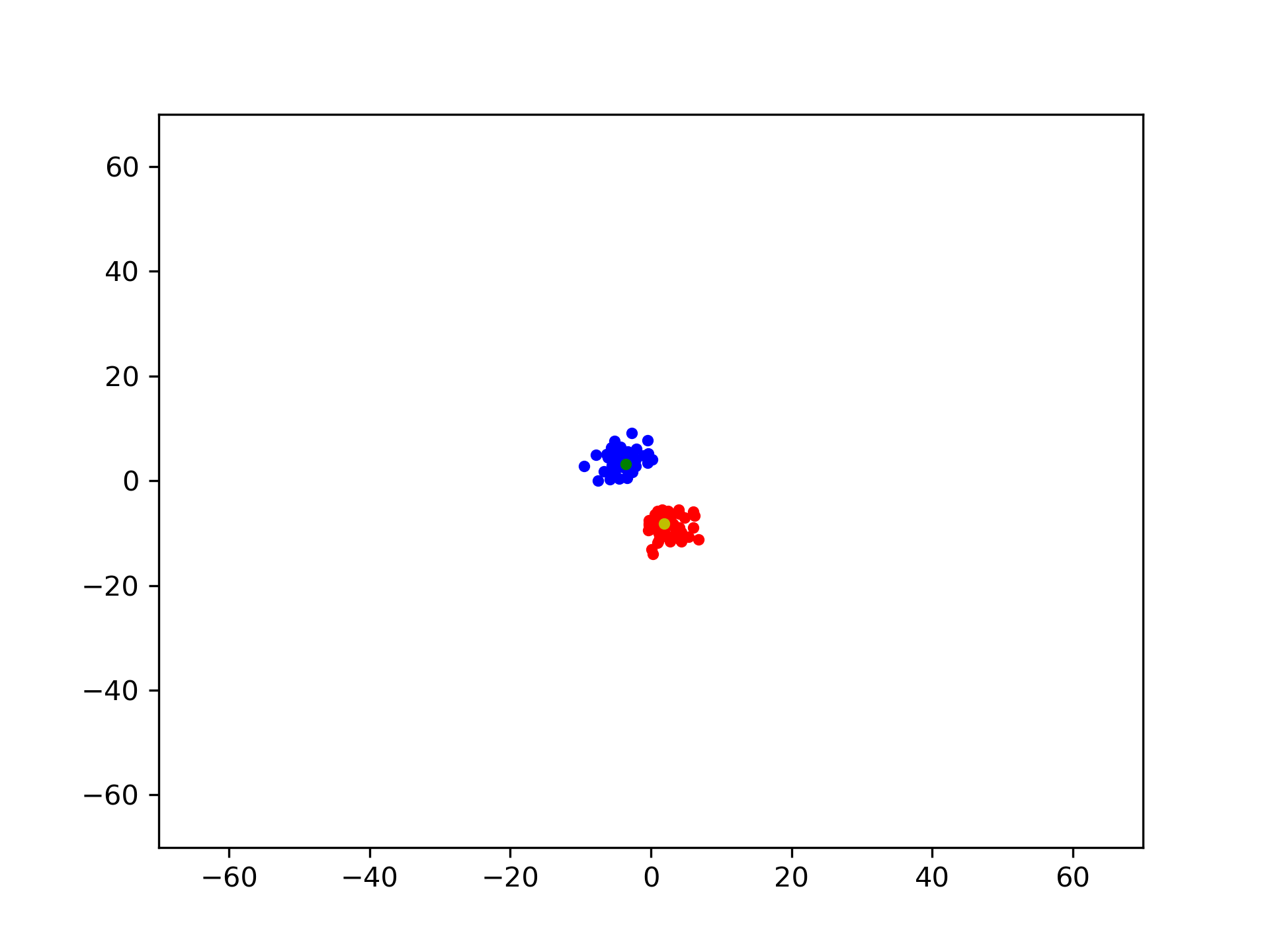}
         \includegraphics[width=0.35\textwidth]{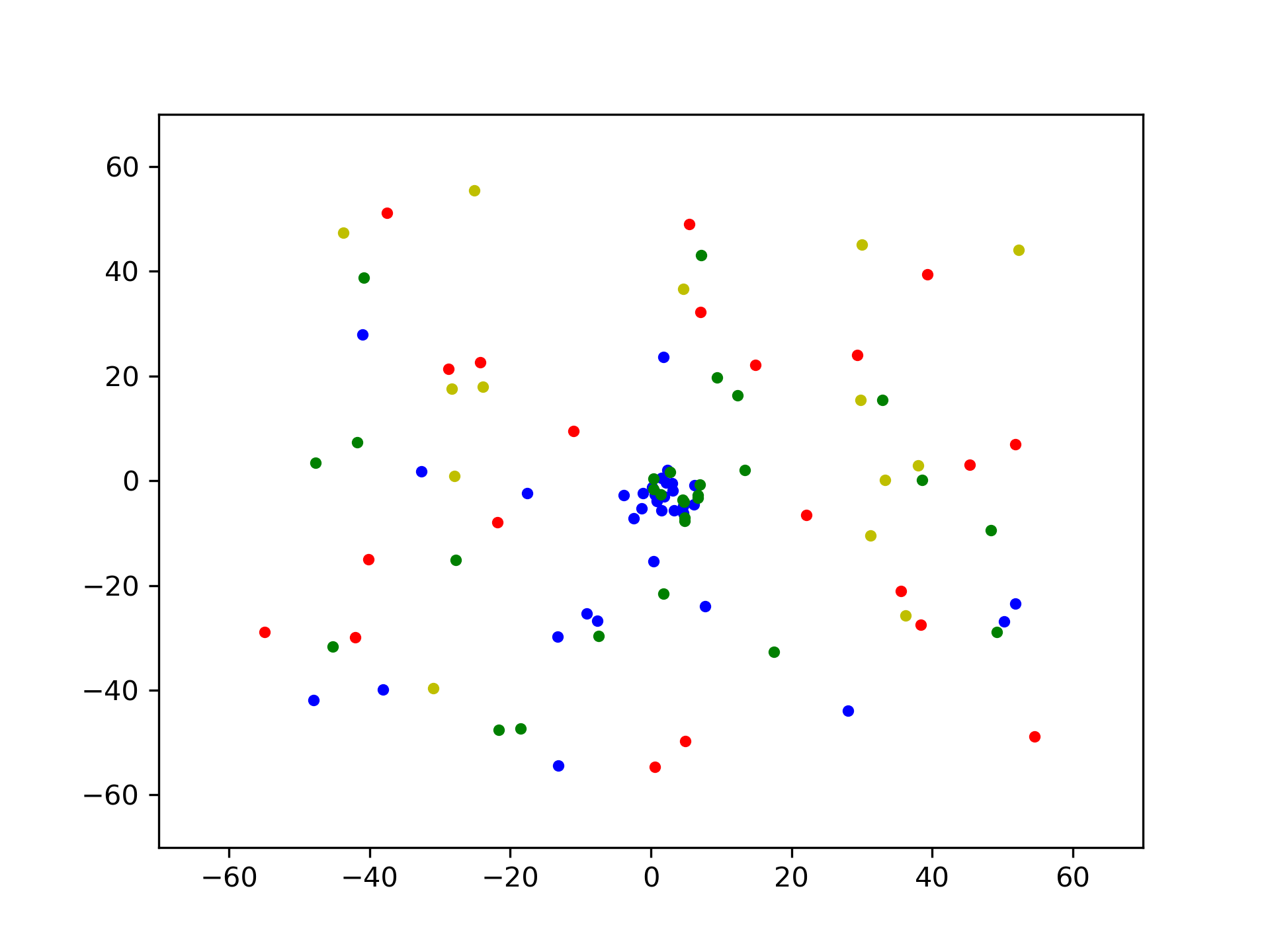}
         \includegraphics[width=0.35\textwidth]{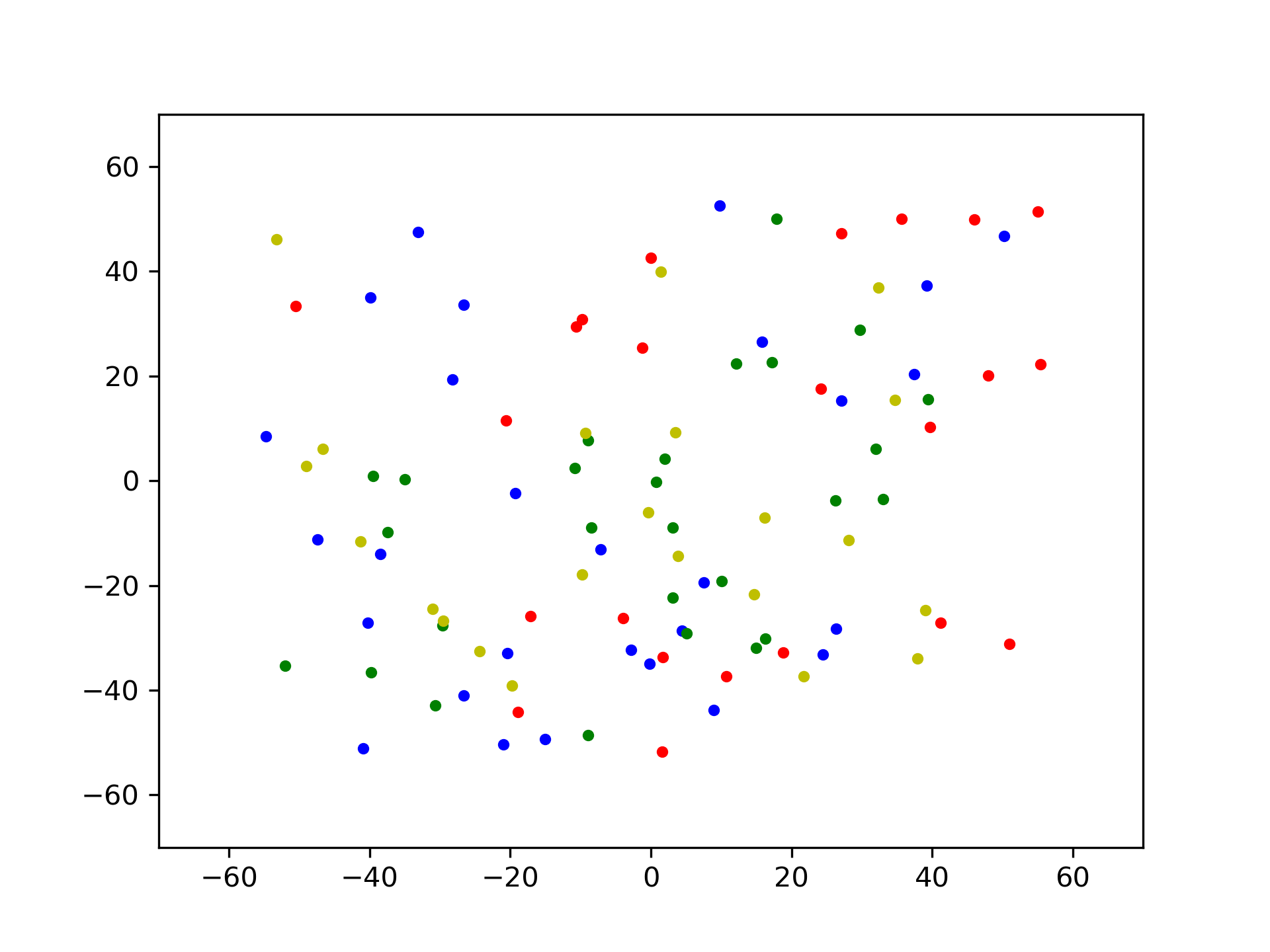}
%
\caption{Dynamics in stable and metastable zones. The left-side image shows a single-cluster; the center shows the main cluster and the right-side image show
the cloud of stable states. Each of these images shows four different configurations of $S_{i}$ and $B_{i}$; where: blue points refer to $S_{i} = 0$ and $B_{i} = 0$; red
points when $S_{i} = 1$ and $B_{i} = 1$; green points when $S_{i} = 0$ and $B_{i} = 1$ and finally yellow points when $S_{i} = 1$ and $B_{i} = 0$.}
\label{fig_ABC}
\end{figure}

As is well known, in a first-order phase transition we normally have two clearly distinct states that can be seen in a caloric curve as in Fig.~\ref{fig_canonical}, in which
we only observe a low-energy and a high-energy state separated by a gap. The nature of these stable states can be understood by thinking in terms of the solid-liquid phase transition, where the particles are together at low energies and without being
disturbed, in a state of thermodynamic equilibrium. On the other hand, at higher energies the particles are separated and move independently, reaching a new state of equilibrium where they do not interact with each other as much, and that is what we call the liquid phase.

However, metastable states are transient states located outside the equilibrium region of the phase diagram, that are not
accesible from the canonical ensemble. A typical example of metastable state it the superheated liquid state which, when
being perturbed, spontaneously reaches the gaseous state. More closely related to our model is the Hamiltonian Mean
Field (HMF) model, which is useful as a model system to study long-range interactions, and which presents metastable states~\cite{Dauxois2002,Atenas2017}.

In order to gain some insight into the behavior of agents in their stable and metastable states, Fig.\ref{fig_ABC} shows three snapshots of the agents configuration at
different energies, $E$=-30.0 (left), $E$=3.0 (center) and $E$=20.0 (right), where in each of the panels a particular collective behavior can be seen. In all
cases the dynamics begins with an initial configuration of two groups separated by a distance larger than $R_c$ so that they do not interact with each other. These
groups are shown as blue ($S = B = 0$) and red ($S = B = 1$) points. It corresponds to the ground state of the system.

In the first zone, at low energies, we have two clusters of agents where each group is mostly consistent with their beliefs, while a few on each cluster only belongs to the cluster by manifesting a false opinion. These correspond to the green points ($S = 0$ and $B = 1$) in the blue group and
the yellow points ($S = 1$ and $B = 0$) in the red group.

In the second zone we have metastability, and in this case some of the agents look for others who have the
same opinion (even if they have to ``fake'' said opinion) and form a main cluster that is, a group of people who interact and talk about a particular topic), while
the rest do not agree and they move away to form a kind of ``cloud''. Finally, at high energies ($E>5$ units) we have a stable state where only the ``cloud'' exists: the agents barely interact and these individuals can be interpreted as being detached from both their own beliefs and the opinion of others.

In both the caloric curves and the configurations of the particles (agents), we see striking similarities with the self-gravitating ring (SGR) model by
Casetti and Nardini~\cite{Casetti2010}, despite the fact that in our model the interactions are short-ranged.

It is common for metastable states to manifest interesting dynamical properties, such as anomalous diffusion~\cite{Klix2010,Sokolov2012}. For instance, the superheated solid constitutes a
metastable state where a \emph{cooperative mobility} appears when the energy is incremented~\cite{Davis2011c,Zhang2013a,OlguinArias2019}. In our case we also expect to have cooperative mobility because of the abrupt nature of the transition between cluster and cloud, but
is probably given according to interactions due to the four different configurations as we have described.

\subsection{Stochastic dynamics and multistability}

\begin{figure}[h!]
         \includegraphics[height=5.2cm]{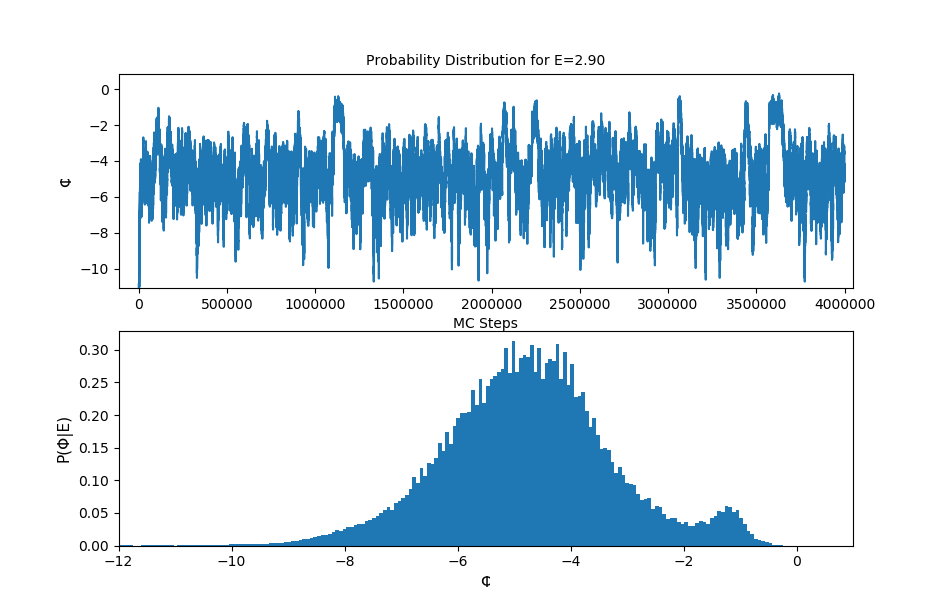}
         \includegraphics[height=5.2cm]{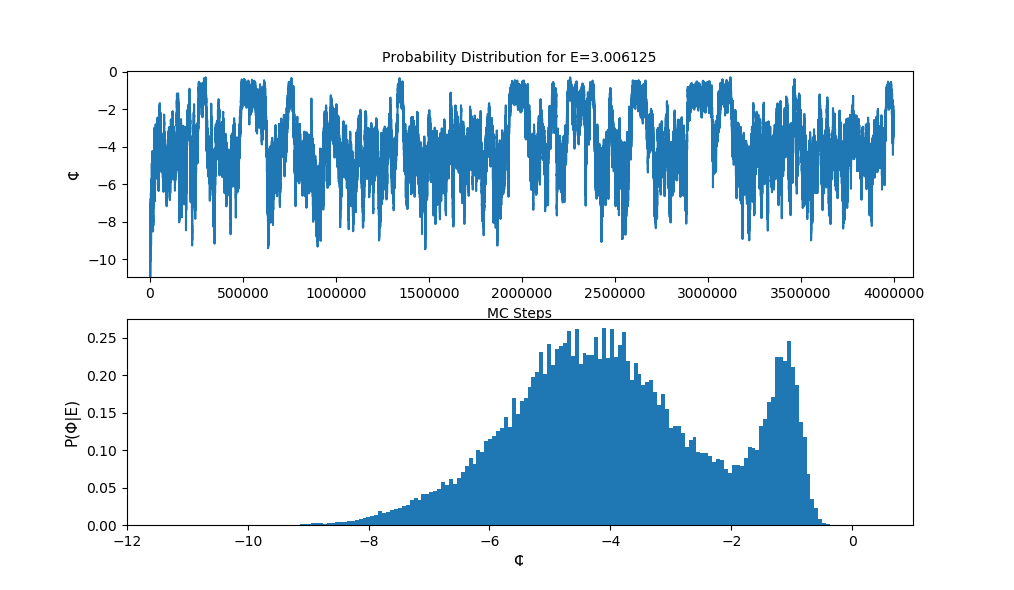}
\caption{Left panel, trace plot (above) of potential energy $\Phi$ during 4 million Monte Carlo steps, and its histogram (below) for $E$=2.90. Right panel, trace plot (above) of potential energy $\Phi$ during 4 million Monte Carlo steps, and its histogram (below) for $E$=3.006125.}
\label{fig_histA}
\end{figure}

\begin{figure}
         \includegraphics[height=5.2cm]{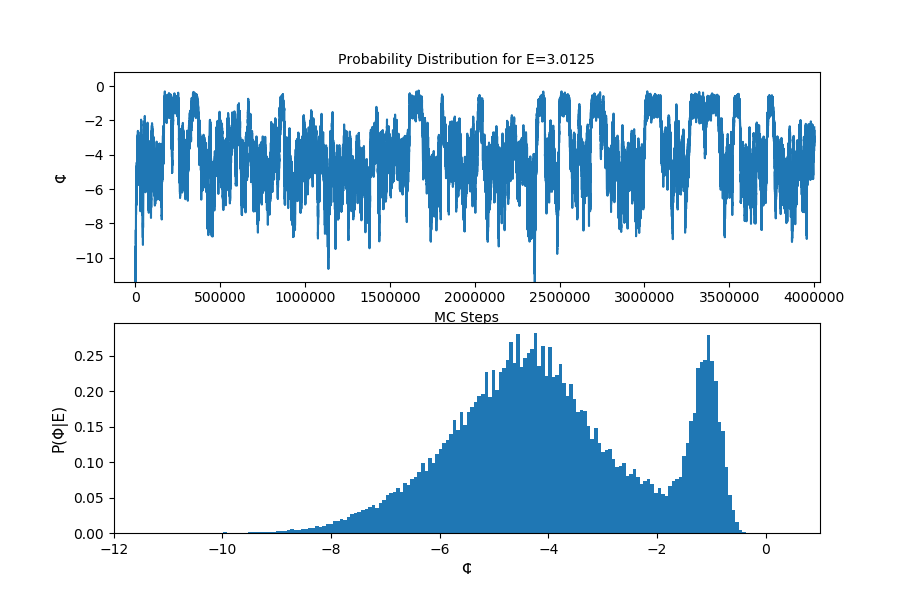}
         \includegraphics[height=5.2cm]{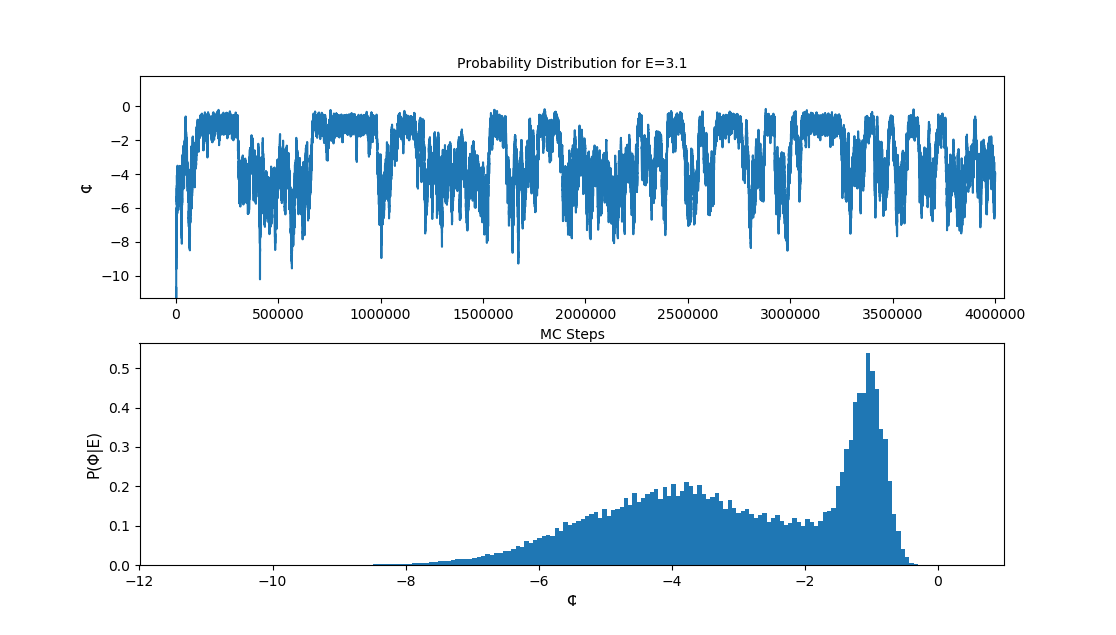}
\caption{Left panel, trace plot (above) of potential energy $\Phi$ during 4 million Monte Carlo steps, and its histogram (below) for $E$=3.0125. Right panel, trace plot (above) of potential energy $\Phi$ during 4 million Monte Carlo steps, and its histogram (below) for $E$=3.1.}
\label{fig_histB}
\end{figure}

We have also explored the stochastic dynamics ocurring in the system in these microcanonical metastable states~\cite{Animation}. Figs.~\ref{fig_histA} and \ref{fig_histB} show a
phenomenon of dynamical multistability near the higher energy transition point at $E$=3, which is similar to the oscillation of phases that ocurrs for small systems near
a first-order phase transition~\cite{Alfe2011, Moreno2018} and have been also reported previously in complex dynamical systems~\cite{Sneppen2012}.

We see in the left panels of Fig.~\ref{fig_histA} a distribution that begins to have a bimodal shape, while most of the distribution is
concentrated in the first peak, which according to Fig.~\ref{fig_microcanonical} at this energy is a stable
configuration. However, in the right panel it is shown that there are more clear signs of a bimodal distribution, and this is due to an increase
in total energy, which favors the higher potential energy state (above -2 units). It is well-known that metastable states present bimodal distributions of total energy (in the canonical ensemble) and potential energy (in the microcanonical ensemble). In the latter case, as a consequence of the shape of the microcanonical caloric curve, indicating a coexistence
of phases~\cite{Eryurek2007}.

In Fig.~\ref{fig_histB}, two other sequential panels are shown with increasing energy and maintaining a bimodal distribution but in these cases more pronounced in the secondary peak;
rather, the distribution shifts according to the fact that the states tend to stay at one of the potential energy minima for a certain time until they spontaneously switch to another.
These metastable states have usually short lifetimes, but can last a considerable proportion of the total simulation time. In the right panel of Fig.~\ref{fig_histB}, we see that for
a higher energy, the secondary peak has the highest density and this means that it has become the most stable state. These simulation results suggest a transition energy at
$E\sim 3.006125$. Additionally, Figs.~\ref{fig_histA} and \ref{fig_histB} also suggest the appearance of multiple metastable states close to one another. The dynamics that are
observed are strongly dependent on the particular initial conditions imposed~\cite{Atenas2017}, since they determine the local minimum the system is closest to at the beginning.


\section{Concluding Remarks}

In summary, we have presented canonical and microcanonical simulations of an off-lattice variant of the Potts model which arises from a maximum entropy inference procedure, where
agents move freely and interact depending of their opinions, under one of four different possible configurations for $Q$=2. Given that $B \in \{0, 1\}$ we have two combinations
with $S = B$, corresponding to an agent manifesting a consistent opinion, and two combinations with $S \neq B$, where the agent expresses an opinion contrary to its internal belief.

This variant of the Potts model presents a first-order phase transition with multiple metastable states which appear on a \textit{V-shaped} region of the caloric curve with negative
specific heat, between $E= -25.6525$ and $E=15.2448$ in Fig.\ref{fig_mix}.We have presented the dynamics of this system in a graphical way in three different zones, namely
``two clusters'', ``main cluster plus cloud'' and ``cloud'', in which the main cluster plus cloud configuration is clearly a metastable state that is forbidden in the canonical ensemble.

The dynamical phenomenon of multistability, being a common phenomenon in complex dynamical systems, is also present in our case as the system explores the metastable states
(local minima) on both sides of a bimodal distribution of potential energy in the \textit{V-shaped} region. These states have short lifetimes, and the system oscillates frequently
between the different local minima. In this work we have presented a system with short-range interactions in which surprisingly there is a
similar behavior to a long-range interacting system. This last result is interesting for future work related to this type of interacting spin models.


\section{Acknowledgements}

SD gratefully acknowledges partial funding from FONDECYT 1171127 and Anillo ACT-172101 grants.


\appendix
\title{Appendix A}
\appendix

\section{Derivation of the model}

In this section we show the mathematical development of the main model used in this work. We start from the constraints
\begin{align}
    \label{eq:A.1}
    P(S_{i}=B_{i}|I) = \Big<\delta(S_{i},B_{i})\Big>_{I} & = p_c \qquad\forall i=1,i\ldots,N \\
    \label{eq:A.2}
    P(S_{i}=S_{j}|r_{i,j}<R_{c},I) = \frac{\Big<\delta(S_{i},S_{j})\Theta(R_{c}-r_{i,j})\Big>_{I}}{\Big<\Theta(R_{c},r_{i,j})\Big>_{I}} & = p_A \qquad\forall i,j=1,\ldots,N
\end{align}
\noindent
The constraint in equation (\ref{eq:A.2}) can be written as
\begin{equation}
    \Big<\Theta(R_{c},r_{i,j})(\delta(S_{i},S_{j})-p_A)\Big>_{I} = 0 \qquad\forall i=1,\ldots,N
\end{equation}
\noindent
Maximizing the Shannon entropy under the constraints in Eqs. \ref{eq:A.1} and \ref{eq:A.2}, the model is
\begin{equation}\label{eq:A.4}
    P(S_{1},...,S_{N}|B_{1},...,B_{N},I)=  \frac{1}{Z}\exp\Big(-\sum_{i=1}^{N}\lambda_i\delta(S_{i},B_{i})-\sum_{i=1}^{N}\sum_{j=1}^{N}\mu_{ij}\Theta(R_{c}-r_{ij})(\delta(S_{i},S_{j}-p_A))
\Big)
\end{equation}
which resembles a canonical ensemble. Therefore, after assuming $\lambda_i=\lambda$ and $\mu_{ij} = \mu$ we can write Eq.~\ref{eq:A.4} as
\begin{equation}
P(S_1,\ldots,S_N|B_1,\ldots,B_N,I) := \frac{1}{Z}\exp(-\beta H(S_1,\ldots,S_N; B_1,\ldots,B_N, \lambda, \mu))
\end{equation}
with the definition of a Hamiltonian
\begin{equation}
  H = -C\sum_{i=1}^{N} \Big<s_{i},B_{i}\Big> -CN -\frac{1}{2}J\sum_{i=1}^{N}\sum_{<j\neq i>} \Big<s_{i},s_{j}\Big> + \frac{R}{2}\sum_{i=1}^{N}\sum_{<j\neq i>}1,
\end{equation}
where $\big<a,b\big> := 2\delta(a,b)-1$ and with the new parameters
\begin{align}
    C & := -\frac{\lambda}{2\beta} \\
    J & := -\frac{\mu}{\beta} \\
    R & := J(2 p_A-1).
\end{align}

The choice $p_A=\frac{1}{2}$ eliminates the overcrowding term, and the constant term $-CN$ is absorbed into the normalization factor, i.e. the partition function.

\end{document}